\documentclass[conference]{IEEEtran}
\IEEEoverridecommandlockouts
\usepackage{cite}
\usepackage{amsmath,amssymb,amsfonts}
\usepackage{algorithmic}
\usepackage{graphicx}
\usepackage{textcomp}
\usepackage{xcolor}

\usepackage{bm}
\usepackage{multicol}
\usepackage{multirow}
\usepackage{makecell}
\usepackage{booktabs}
\usepackage{algorithm}
\usepackage{algorithmic}
\usepackage{subfigure}

\newcommand{\bmmc}[1]{\bm{\mathcal{#1}}}
\DeclareMathOperator*{\argmin}{argmin} 

\makeatletter
\newcommand{\linebreakand}{%
  \end{@IEEEauthorhalign}
  \hfill\mbox{}\par
  \mbox{}\hfill\begin{@IEEEauthorhalign}
}
\makeatother

\def\BibTeX{{\rm B\kern-.05em{\sc i\kern-.025em b}\kern-.08em
    T\kern-.1667em\lower.7ex\hbox{E}\kern-.125emX}}
\begin{document}

\title{In-Sensor Radio Frequency Computing for Energy-Efficient Intelligent Radar 
}

\author{\IEEEauthorblockN{Yang Sui}
\IEEEauthorblockA{
\textit{Rutgers University}\\
yang.sui@rutgers.edu}
\and
\IEEEauthorblockN{Minning Zhu}
\IEEEauthorblockA{
\textit{Rutgers University}\\
minning.zhu@rutgers.edu }
\and
\IEEEauthorblockN{Lingyi Huang}
\IEEEauthorblockA{
\textit{Rutgers University}\\
lingyi.huang@rutgers.edu} 
\linebreakand
\IEEEauthorblockN{Chung-Tse Michael Wu}
\IEEEauthorblockA{
\textit{Rutgers University}\\
ctm.wu@rutgers.edu}
\and
\IEEEauthorblockN{Bo Yuan}
\IEEEauthorblockA{
\textit{Rutgers University}\\
bo.yuan@soe.rutgers.edu}
}

\maketitle

\begin{abstract}
Radio Frequency Neural Networks (RFNNs) have demonstrated advantages in realizing intelligent applications across various domains. However, as the model size of deep neural networks rapidly increases, implementing large-scale RFNN in practice requires an extensive number of RF interferometers and consumes a substantial amount of energy. To address this challenge, we propose to utilize low-rank decomposition to transform a large-scale RFNN into a compact RFNN while almost preserving its accuracy. Specifically, we develop a Tensor-Train RFNN (TT-RFNN) where each layer comprises a sequence of low-rank third-order tensors, leading to a notable reduction in parameter count, thereby optimizing RF interferometer utilization in comparison to the original large-scale RFNN. Additionally, considering the inherent physical errors when mapping TT-RFNN to RF device parameters in real-world deployment, from a general perspective, we construct the Robust TT-RFNN (RTT-RFNN) by incorporating a robustness solver on TT-RFNN to enhance its robustness. To adapt the RTT-RFNN to varying requirements of reshaping operations, we further provide a reconfigurable reshaping solution employing RF switch matrices. Empirical evaluations conducted on MNIST and CIFAR-10 datasets show the effectiveness of our proposed method.

\end{abstract}


\section{Introduction}

Radar, as the fundamental and important remote sensing system, has served as the backbone sensing solution in a variety of scientific and engineering applications, such as weather prediction, cosmic exploration, and autonomous navigation. By transmitting the radio waves and receiving the reflected waves, radar can determine the distance, angle, and/or radical velocity of the stationary or moving objects. 

Traditionally, a radar system consists of a sensing module that generates, transmits, and receives electromagnetic waves, and a computing module that processes and analyzes the physical signals. By adopting advanced signal processing and machine learning algorithms, the state-of-the-art computing modules of the radar systems can extract and perceive richer high-level characteristics of the target object, e.g., the shape and category.

From the perspective of hardware design, the split sensing and computing modules bring insufficient hardware efficiency. More specifically, the use of costly ADCs connecting the RF-domain sensor and digital-domain processor not only brings significant data movement but also high power consumption. To address this challenge, very recently some works have proposed to perform computing in the RF domain, i.e., in-sensor RF computing. In particular, our prior work RFNN \cite{10186076}, proposes to move the execution of neural networks, which are typically implemented in the back-end digital processor, to the front-end RF sensor. By leveraging the characteristic that the 2$\times$2 RF interferometer, a basic computing unit that the RF circuit can maturely implement, can be used to build arbitrary matrices, the topology of a feedforward neural network, in principle, can be conveniently mapped on the RF circuit, enabling the RF-domain neural network computing. Notice that though such Mach-Zehnder Interferometer (MZI)-like mapping is also used in optical computing, the required unit interferometer cell usually goes up to 100$\times$ of its wavelength, while in the RF domain, the unit cell is of sub-wavelength size. Moreover, optic-electric/electric-optic conversion, as a costly overhead component in the system, can be completely removed when performing the computation in the RF domain, implying the promising potential of in-sensor RF computing.

In this work, we investigate the feasibility of RFNN from a more practical perspective. More specifically, considering the neural network models deployed in real-world applications typically consist of multiple layers with hundreds of even thousands of neurons per layer, the original RFNN designed for very small neural networks, due to the physical constraints of RF circuit size, cannot be readily used in practical applications. To address this challenge, we propose to apply low-rank tensor train (TT) decomposition \cite{yin2021towards, yin2022hodec, gong2023ette}, an effective model compression technique \cite{han2015learning, lin2020hrank, sui2021chip}, to develop a compact RFNN with negligible accuracy loss. The resulting more efficient design, namely TT-RFNN, can notably reduce the number of the required computational RF interferometers and hence the overall power consumption of entire RF circuits.  

Moreover, considering mapping the neural network models on RF devices would inevitably introduce physical errors, which may be incurred by various noise sources such as phase bias and device variations, we further develop a robust solution, namely RTT-RFNN, to ensure that the obtained TT-RFNN models can still achieve the desired task performance when deployed in the practical RF devices with physical perturbation. To that end, we formulate our design goal as a constrained optimization problem and then solve it by using the alternating direction optimization method to obtain a feasible solution.



In addition, to support the different tensor reshaping, an important operation in the proposed RTT-RFNN, and provide sufficient architectural flexibility, we propose a reconfigurable hardware architecture that utilizes RF switches matrix as the interconnect layer \cite{daneshmand2011rf}, thereby enabling the reconfigurable interconnect mesh in the circuit structures. Overall, our contributions are summarized as follows: 
\begin{itemize}
    \item We propose to use tensor-train factorization to decompose a large-scale RFNN into a series of compact low-rank small tensor cores, significantly reducing the number of required RF interferometers and facilitating the deployment of large-scale networks on RF device parameters.

    \item We formulate the design of robust TT-RFNN as an optimization problem, and solve it by using an alternating direction optimization approach, consisting of a robust training process with unitary constraints, enhancing the robustness of TT-RFNN against physical noises. 

    \item We propose switch matrices-based reconfigurable reshaping operation tailored for the RTT-RFNN, improving the implementation flexibility of RTT-RFNN hardware. 

    \item Experimental results on different datasets and neural network models show our proposed solution can achieve promising hardware performance with good efficiency and robustness.

\end{itemize}

\section{Preliminaries}

We denote the order $d$ tensor as $\bm{\mathcal{X}}\in\mathbb{R}^{n_1 \times n_2 \times \cdots \times n_d}$, the matrix as $\bm{X}\in\mathbb{R}^{m \times n}$ and the vector as $\bm{x}\in\mathbb{R}^{n}$, respectively. Let $\bm{\mathcal{X}}_{(i_1,\cdots,i_d)}$ and $\bm{X}_{(i, j)}$ denote the individual element at the position $(i_1,\cdots,i_d)$ of tensor $\bm{\mathcal{X}}$, and $(i, j)$ of matrix $\bm{X}$, respectively, where $i_1 \le n_1, i_2 \le n_2, \cdots, i_d \le n_d$.

\subsection{RF Interferometer}


\begin{figure}[t] 
\centering
 \subfigure[]
     {\includegraphics[width=0.48\textwidth]{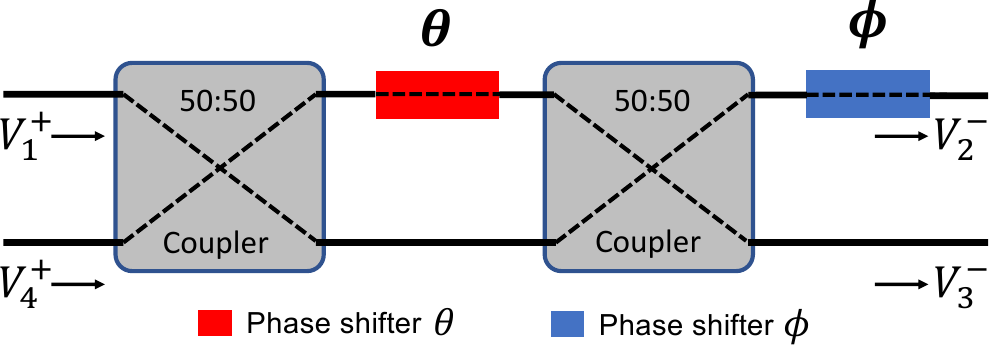}
     \label{fig:RF_unit1}
     } \\
          \vspace{-2mm}
 \subfigure[]
 {\includegraphics[width=0.48\textwidth]{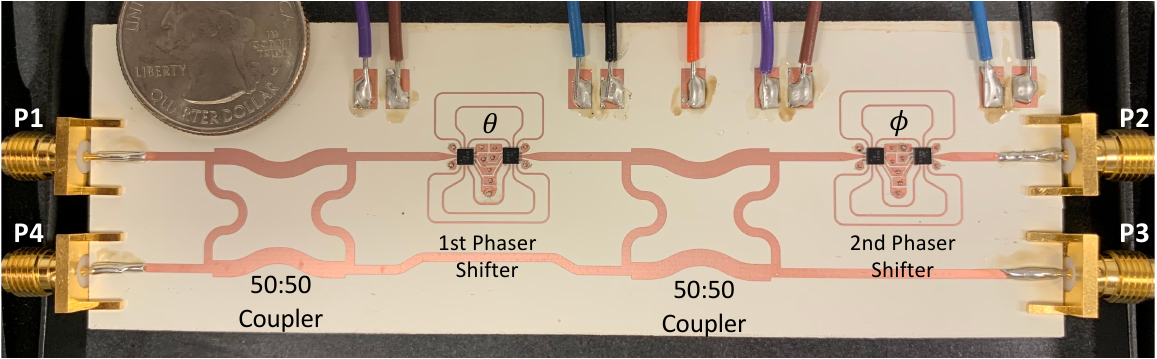}
     \label{fig:RF_unit2}
     } 
     \caption{RF interferometer cell (a) schematic and (b) device photo \cite{10186076}.}
    \label{fig:RF_unit}
    \vspace{-2mm}
\end{figure}

\begin{figure}[t] 
\centering
 \subfigure[Triangular mesh style.]
     {\includegraphics[width=0.48\textwidth]{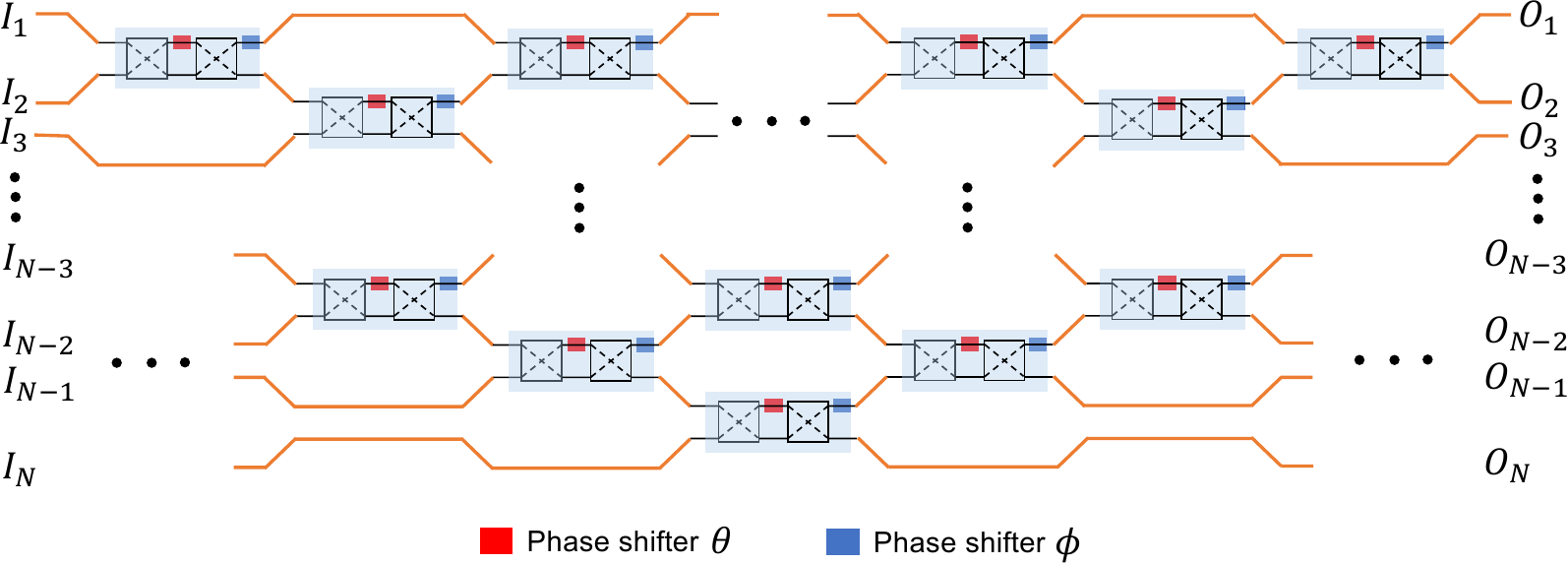}
     \label{fig:Triangular}
     } \\
 \subfigure[Rectangular mesh style.]
 {\includegraphics[width=0.48\textwidth]{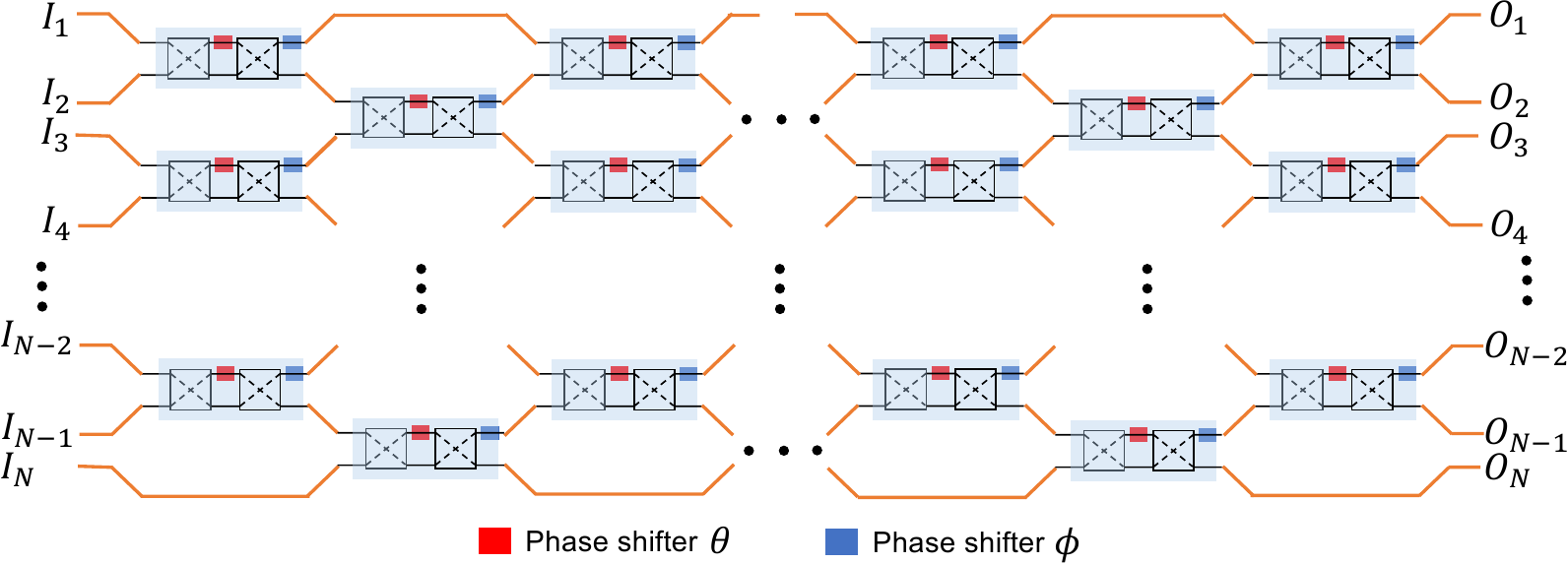}
     \label{fig:Rectangular}
     } 
     \caption{RFNN in synthesizing an ${N}$ dimension unitary matrix with (a) triangular and (b) rectangular mesh style.}
    \label{fig:tri_rec}
    \vspace{-2mm}
\end{figure}

Recently, a linear RF interferometer made of two 50:50 quadrature hybrid couplers and two phase shifters for realizing reconfigurable microwave artificial neural network is proposed by \cite{10186076}, whereas its structure is illustrated in Fig. \ref{fig:RF_unit}. The total voltage transformation matrix of the device can be expressed as:
\begin{equation}
\begin{aligned}
    \begin{bmatrix} 
    V_2^-\\ 
    V_3^-
    \end{bmatrix} = &\bm{T}\begin{bmatrix}
                            V_1^+\\
                            V_4^+
                            \end{bmatrix} \\
                    = &je^{-j\frac{\theta}{2}}
                    \begin{bmatrix}
                        e^{-j\phi}\sin\left(\frac{\theta}{2}\right) & e^{-j\phi}\cos\left(\frac{\theta}{2}\right)\\   \cos\left(\frac{\theta}{2}\right) & -\sin\left(\frac{\theta}{2}\right)
                    \end{bmatrix}
                    \begin{bmatrix}
                        V_1^+ \\ 
                        V_4^+
                    \end{bmatrix},
\end{aligned}
\label{eqn:voltage_matrix}
\end{equation}
where, ${V_1^+}$ and ${V_4^+}$ represent the inward propagating voltage magnitude at the input ports and ${V_2^-}$ and ${V_3^-}$ represent the outward propagating voltage magnitude at the output ports. In general, this device can be utilized to map arbitrary unitary matrix ${\bm{U} \in \mathbb{R}^{N\times N}}$ (triangular style):
\begin{equation}
\begin{aligned}
\bm{U}=\bm{D} \cdot \bm{R}^{\left(S\right)}\cdot \bm{R}^{\left(S-1\right)} \cdots \bm{R}^{\left(1\right)},
\end{aligned}
\label{eq:udrrr}
\end{equation}
where, ${S=N(N-1)/2}$ and  ${\bm{D} \in \mathbb{R}^{N \times N}}$ is a diagonal matrix, the module of which is an identity matrix. ${\bm{R}^{\left(s\right)} \in \mathbb{R}^{{N\times N}}}$ represents an $N$-dimensional rotational matrix formed from an identity matrix with four adjacent elements on the diagonal replaced by the transformation matrix. The indices of the replaced four elements are determined by the location of the device among the ${N}$ propagation channels. For example, if the device crosses channel ${p}$ and ${q}$, where ${q = p+1}$, then the elements ${r_{pp}}$, ${r_{pq}}$,\ \ ${r_{qp}}$,\ and\ \ ${r_{qq}}$\ of ${\bm{R}^{\left(s\right)}}$ are replaced with the device transformation matrix $\bm{T}$, as follows:
\begin{equation}
\begin{aligned}
\bm{R}^{\left(s\right)}=\left[\begin{matrix}\begin{matrix}1&\cdots&0\\\vdots&\ddots&\vdots\\0&\cdots&r_{pp}\\\end{matrix}&\begin{matrix}0&\cdots&0\\\vdots&\ddots&\vdots\\r_{pq}&\cdots&0\\\end{matrix}\\\begin{matrix}0&\cdots&r_{qp}\\\vdots&\ddots&\vdots\\0&\cdots&0\\\end{matrix}&\begin{matrix}r_{qq}&\cdots&0\\\vdots&\ddots&\vdots\\0&\cdots&1\\\end{matrix}\\\end{matrix}\right],
\end{aligned}
\label{eq:rs}
\end{equation}
\begin{equation}
\begin{aligned}
\left[\begin{matrix}r_{pp}&r_{pq}\\r_{qp}&r_{qq}\\\end{matrix}\right]=\bm{T}.
\end{aligned}
\label{eq:RtoT}
\end{equation}
Note that such decomposition can be realized following a triangular style or a rectangular style (shown in Fig. \ref{fig:Triangular} and Fig. \ref{fig:Rectangular}). Both require $S$ devices through $N$ propagation channels ($N$ is also the number of inputs/outputs).



Because an arbitrary weight matrix ${\bm{W}\in\mathbb{R}^{N\times N}}$ can be decomposed to unitary matrices through the singular value decomposition (SVD) as ${\bm{W}=\bm{U} \bm{\Sigma} \bm{V}^\ast}$, this general property enables the mapping of non-unitary matrices on the RF devices. It is worth noting that, instead of decomposing ${\bm{U}}$, we decomposed ${\bm{U}^\ast}$, so that the leftover diagonal matrix ${\bm{D}_{U^\ast}}$ can also be absorbed into ${\bm{\Sigma}}$, where a new diagonal matrix ${\bm{\Sigma}^\prime=\bm{D^\ast}_{U^\ast} \bm{\Sigma} \bm{D}_{V^\ast}}$. Therefore, the total number of devices used for mapping the weight matrix $\bm{W}$ remains to be ${N^2}$.

\subsection{Tensor-Train Decomposition}

\begin{figure}[t] 
    \centering 
    \includegraphics[width=1\linewidth]{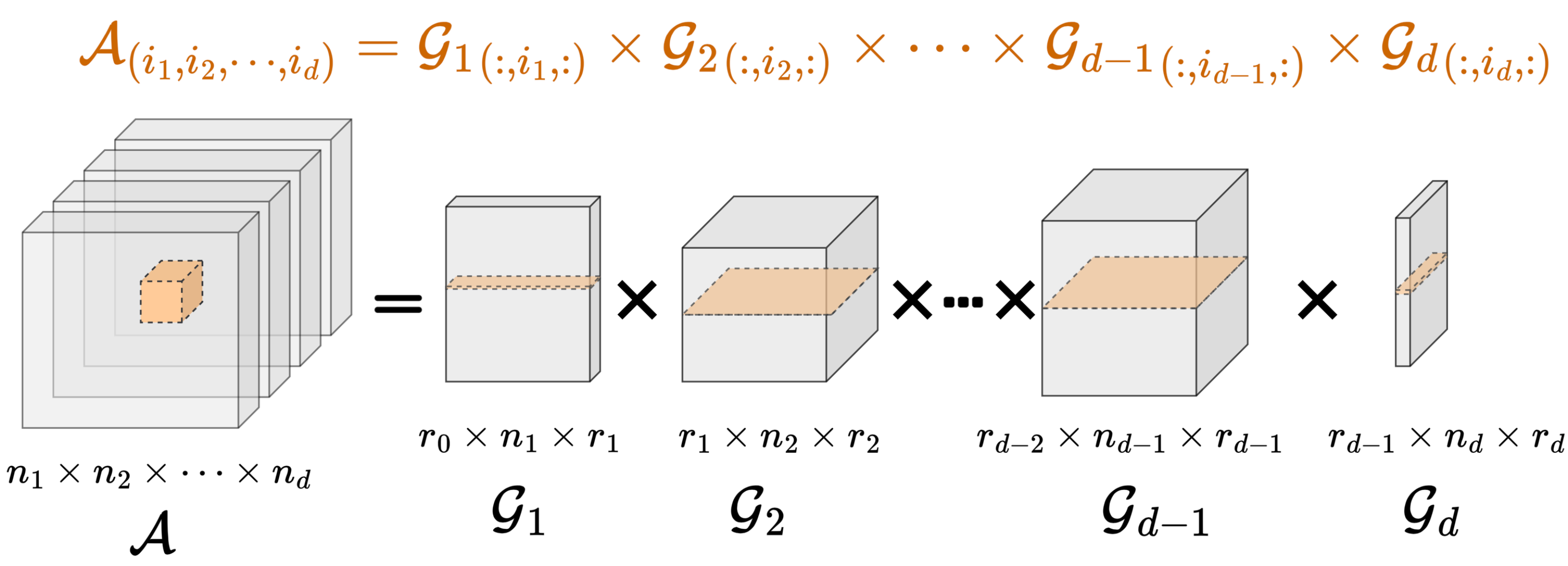}
\caption{Illustration of tensor-train decomposition.}
\label{fig:ttd}
\vspace{-2mm}
\end{figure}

Given a high-order tensor $\bm{\mathcal{A}}\in\mathbb{R}^{n_1\times n_2\times\cdots\times n_d}$, it can be decomposed using the Tensor-Train (TT) decomposition into a series of third-order tensors. This representation is referred to as the TT-format for the tensor $\bm{\mathcal{A}}$, as follows:
\begin{equation}
\begin{aligned}
    &\bm{\mathcal{A}}_{(i_1,i_2,\cdots,i_d)}={\bm{\mathcal{G}}_1}_{(:,i_1,:)}{\bm{\mathcal{G}}_2}_{(:,i_2,:)}\cdots{\bm{\mathcal{G}}_d}_{(:,i_d,:)}\\
    &=\sum_{\alpha_0,\alpha_1\cdots\alpha_d}^{r_0,r_1,\cdots r_d}{\bm{\mathcal{G}}_1}_{(\alpha_0,i_1,\alpha_1)}{\bm{\mathcal{G}}_2}_{(\alpha_1,i_2,\alpha_2)}\cdots
    {\bm{\mathcal{G}}_d}_{(\alpha_{d-1},i_d,\alpha_d)},
\end{aligned}
\label{eq1}
\end{equation}
where $\bm{\mathcal{G}}_k\in\mathbb{R}^{r_{k-1}\times n_k\times r_k}$ are termed as TT-cores for $k=1,2,\cdots,d$. The TT-ranks are represented by ${r_0, r_1, \cdots, r_d}$ with the “boundary conditions” $r_0=r_d=1$. The settings of these ranks significantly determine the efficiency and storage complexity of the TT-format tensor. The maximal TT-rank is defined as $r=\max \{r_1, r_2, \cdots, r_d\}$. Theoretically, TT-format reduces the storage complexity from $\mathcal{O}(n^d)$ to $\mathcal{O}(dnr^2)$. An example is shown in Fig. \ref{fig:ttd}.

\section{Methods}

\subsection{Efficient TT-RFNN}
Consider the calculation of $\bm{y}=\bm{W}\bm{x}$ which includes the weight matrix $\bm{W}\in\mathbb{R}^{O \times I}$ , the input vector $\bm{x}\in\mathbb{R}^I$, and the output $\bm{y}\in\mathbb{R}^{O}$. Here, $O$ and $I$ can be represented as $O=\prod_{k=1}^{d}i_k$ and $I=\prod_{k=1}^{d}j_k$, respectively. We first tensorize the weight matrix $\bm{W}$ to a high-order weight tensor $\bm{\mathcal{W}}\in\mathbb{R}^{i_1\times i_2 \times \cdots \times i_d \times j_1\times j_2\times \cdots \times j_d}$ by the reshaping operation. Then, $\bm{\mathcal{W}}$ can be decomposed to the TT-format:
\begin{equation}
\begin{aligned}
\bm{\mathcal{W}}_{(m_1,m_2,\cdots,m_d,n_1,n_2,\cdots,n_d)} = &{\bm{\mathcal{G}}_1}_{(:,m_1,:)}\cdots {\bm{\mathcal{G}}_d}_{(:,m_d,:)} \\ 
&{\bm{\mathcal{G}}_{d+1}}_{(:,n_1,:)} \cdots{\bm{\mathcal{G}}_{2d}}_{(:,n_d,:)},
\label{eq2}
\end{aligned}
\end{equation}
where each TT-core $\bm{\mathcal{G}}_k\in\mathbb{R}^{r_{k-1}\times i_k\times r_k}$ or $\bm{\mathcal{G}}_k\in\mathbb{R}^{r_{k-1}\times j_{k-d}\times r_k}$ is a third-order tensor. Consequently, the forward propagation represented in the tensor format is as follows:
\begin{equation}
\begin{aligned}
\bm{\mathcal{Y}}_{(m_1,\cdots,m_d)}=\sum_{n_1,\cdots,n_d} &{\bm{\mathcal{G}}_1}_{(:,m_1,:)}\cdots {\bm{\mathcal{G}}_d}_{(:,m_d,:)} \\ 
&{\bm{\mathcal{G}}_{d+1}}_{(:,n_1,:)} \cdots{\bm{\mathcal{G}}_{2d}}_{(:,n_d,:)} \bm{\mathcal{X}}_{(n_1,\cdots,n_d)},
\end{aligned}
\label{eq3}
\end{equation}
where $\bm{\mathcal{X}}\in\mathbb{R}^{j_1\times\cdots\times j_d}$ and $\bm{\mathcal{Y}}\in\mathbb{R}^{i_1\times\cdots\times i_d}$ are the tensorized input and output corresponding to $\bm{x}$ and $\bm{y}$, respectively. 

\begin{figure*}[t] 
	\centering 
	\includegraphics[width=0.9\linewidth]{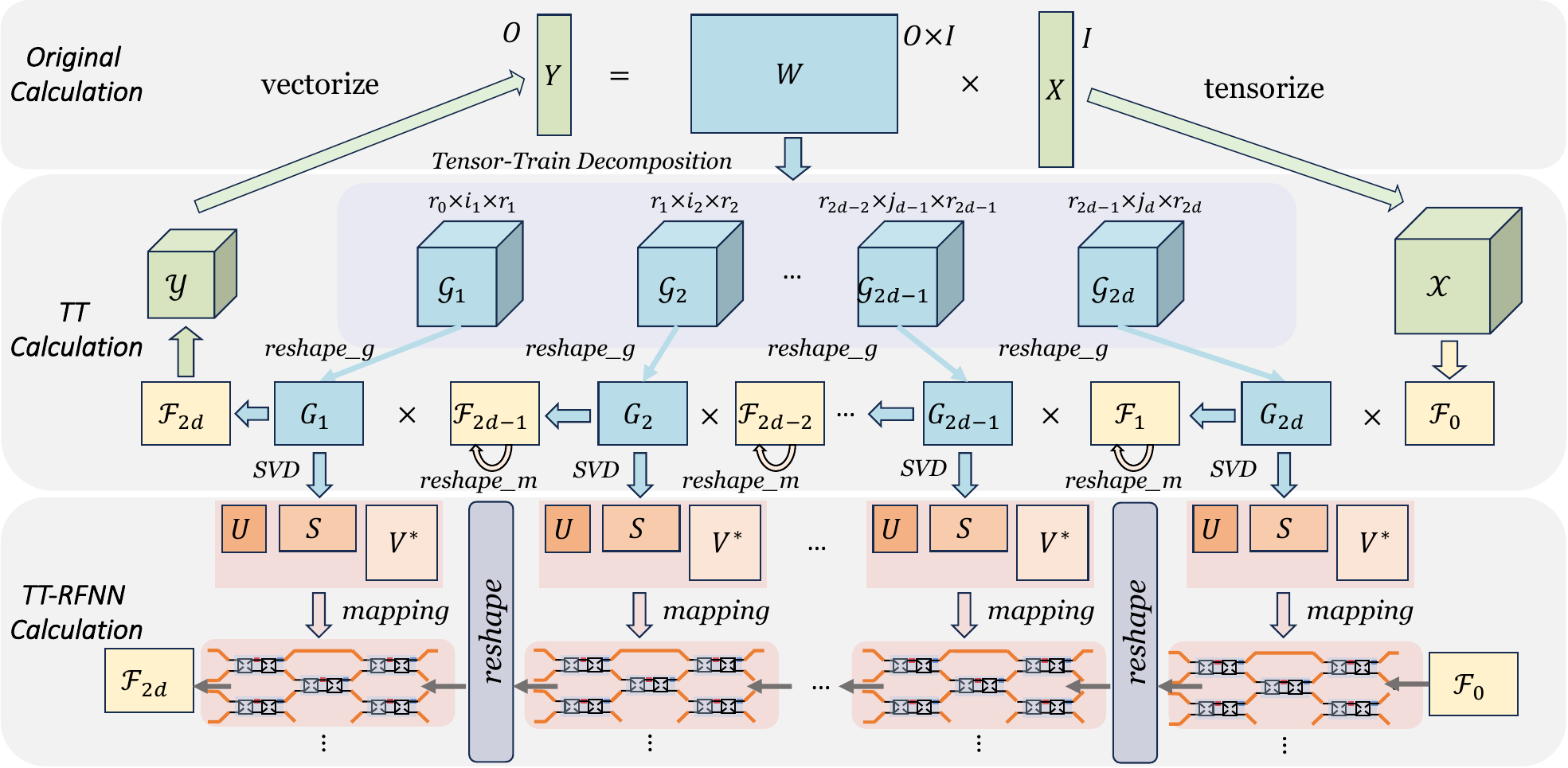}
	\caption{Illustration of TT-RFNN. Top: the original calculation of $\bm{y} = \bm{W} \bm{x}$ where $\bm{W} \in \mathbb{R}^{O \times I}$. Middle: tensor-train decomposition on the original weight $\bm{W}$. Bottom: TT-RFNN calculation. Tensor cores are decomposed by SVD and then mapped to the RF device parameters.}
	\label{fig:ttrfnn}
 \vspace{-2mm}
\end{figure*}

In the case of TT-RFNN, one needs to take into account the reshaping operation, as it plays a significant role in practical circuit implementations (see Fig. \ref{fig:ttrfnn}). Here, we provide a detailed analysis to better the reshaping operation. Notice that $r_0$ to $r_d$ are with $\bm{i}$ and $r_d$ to $r_{2d}$ are with $\bm{j}$. Also, $\bmmc{G}_1$ to $\bmmc{G}_d$ are with $\bm{i}$ and $\bmmc{G}_{d+1}$ to $\bmmc{G}_{2d}$ are with $\bm{j}$.

The first intermediate tensor $\bm{\mathcal{F}}_1$ is calculated from $\texttt{reshape}(f(\bm{\mathcal{G}}_{2d}, \bm{\mathcal{X}}))$. Here $f(\bm{\mathcal{G}}_{2d}, \bm{\mathcal{X}})$ denotes $\bm{G}_{2d}\bm{X}$, where $\bm{G}_{2d} \in \mathbb{R}^{(r_{2d-1}) \times (j_d \times r_{2d})}$ and $\bm{X} \in \mathbb{R}^{(j_d \times r_{2d}) \times (j_1 \times \cdots \times j_{d-1})}$ with $r_{2d} = 1$, are matrices with the reshaping operation from $\bm{\mathcal{G}}_{2d}$ and $ \bm{\mathcal{X}}$, respectively. 

In general, from $l=2d$ to $l=1$ with $2d-l=0$ to $2d-l=2d-1$, the intermediate tensors are calculated as: 
\begin{equation}
\begin{aligned}
\bm{G}_l = \texttt{reshape\_g}(\bm{\mathcal{G}}_l),
\end{aligned}
\label{eqn:reshapeg}
\end{equation}
\begin{equation}
\begin{aligned}
\bm{F}_{2d-l} = \texttt{reshape\_f}(\bm{\mathcal{F}}_{2d-l}),
\end{aligned}
\label{eqn:reshapef}
\end{equation}
\begin{equation}
\begin{aligned}
\bm{\mathcal{F}}_{2d-l+1} = \texttt{reshape\_t}(\bm{G}_l \bm{F}_{2d-l}).
\end{aligned}
\label{eqn:reshapet}
\end{equation}
More specifically, in Eq. \ref{eqn:reshapeg}, $\texttt{reshape\_g}(\cdot)$ transforms the $\bm{\mathcal{G}}_l$ into the matrix $\bm{G}_l \in \mathbb{R}^{(r_{l-1}) \times (j_{l-d} \times r_{l})}$ when $l > d$, or $\bm{G}_l \in \mathbb{R}^{(r_{l-1} \times i_l) \times (r_{l}) }$ when $l \le d$. In Eq. \ref{eqn:reshapef},  $\texttt{reshape\_f}(\cdot)$ converts the $\bm{\mathcal{F}}_{2d-l}$ into the matrix $\bm{F}_{2d-l} \in \mathbb{R}^{(j_{l-d} \times r_{l}) \times (j_{1} \times \cdots \times j_{l-d-1}) }$ when $l > d$ and $\bm{F}_{2d-l} \in \mathbb{R}^{(r_{l}) \times (i_{l+1} \times \cdots \times i_{d}) }$ when $l \le d$. In Eq. \ref{eqn:reshapet}, $\texttt{reshape\_t}(\cdot)$ transforms the matrix to the tensor format with the reshaping operation, from $\bm{F}_{2d-l+1} \in \mathbb{R}^{(r_{l-1}) \times (j_{1} \times \cdots \times j_{l-d-1}) }$ to $\bm{\mathcal{F}}_{2d-l+1} \in \mathbb{R}^{j_{l-d-1} \times r_{l-1} \times  j_{1} \times \cdots \times j_{l-d-2} }$ when $l > d$ and from $\bm{F}_{2d-l+1} \in \mathbb{R}^{ (r_{l-1} \times i_{l}) \times (i_{l+1} \times \cdots \times i_{d})}$ to $\bm{\mathcal{F}}_{2d-l+1} \in \mathbb{R}^{r_{l-1} \times i_{l} \times \cdots \times i_{d}}$ when $l \le d$.

For simplicity, we can pre-reshape $\bmmc{G}$ before the calculation and map them to the RF circuits, then fix it. The intermediate tensor can be directly reshaped from the output of the current layer to the input of the next layer. This can be achieved when calculating the intermediate tensor without the input and output neurons, by using a new reshaping operation $\texttt{reshape\_m}(\cdot)$. Thus, we can omit the $\texttt{reshape\_t}(\cdot)$ as follows: 
\begin{equation}
\begin{aligned}
\bm{G}_l = &\texttt{reshape\_g}(\bm{\mathcal{G}}_l) \\
\bm{{F}}_{2d-l+1} = &\texttt{reshape\_m}(\bm{G}_l\bm{F}_{2d-l})
\end{aligned}
\label{eqn:reshapem}
\end{equation}

In Eq. \ref{eqn:reshapem}, $\texttt{reshape\_m}(\cdot)$ transforms the output matrix to the next input matrix from $(\bm{G}_l\bm{F}_{2d-l}) \in \mathbb{R}^{(r_{l-1}) \times (j_{1} \times j_{2} \times \cdots \times j_{l-d-1}) }$ to $\bm{F}_{2d-l+1} \in \mathbb{R}^{(j_{l-d-1} \times r_{l-1}) \times (j_{1} \times \cdots \times j_{l-d-2})  }$ when $l > d$, and from $(\bm{G}_l\bm{F}_{2d-l}) \in \mathbb{R}^{ (r_{l-1} \times i_{l}) \times (i_{l+1} \times \cdots \times i_{d}) }$ to $\bm{F}_{2d-l+1} \in \mathbb{R}^{ (r_{l-1}) \times (i_{l} \times \cdots \times i_{d}) }$ when $l \le d$.



\subsection{Robustness Solver}

Mapping TT-cores onto RF device parameters would inevitably introduce physical errors. These errors encompass phase bias, device variations, and other noises. As a result, the desired weights are perturbed by random noise. From a general perspective, we characterize all these noises as a random variable following a Gaussian distribution. More specficially, given the matrix format of TT-cores $\bm{G}$, the noise TT-cores in matrix format are represented as $\widehat{\bm{G}}=\widehat{\bm{U}}\widehat{\bm{S}}\widehat{\bm{V}^*}$, where $\bm{U},\bm{S},\bm{V^*} = \texttt{SVD}(\bm{G})$ and $\widehat{\bm{U}} = \bm{U} + \lambda\mathcal{N}(0, I)$, $\widehat{\bm{V^*}} = \bm{V^*} + \lambda\mathcal{N}(0, I)$, representing the practical weight mapping with the perturbed Gaussian noise. Notice that in this paper we only consider the noisy $\widehat{\bm{U}}, \widehat{\bm{V}}$ since they take up almost all parameters. 

Considering that physical noises can lead to performance degradation, it is crucial to improve the robustness of the TT-RFNN and enhance its resistance to such noise. To address this challenge, we propose to use robust training, which adds weights at different iterations with various random noises, thereby developing RTT-RFNN to enhance the performance of TT-RFNN against random disturbances. Also, it is worth noting that when deployed into the RF device parameters, $\widehat{\bm{U}}, \widehat{\bm{V}}$ deviate from being unitary matrices due to noise perturbations. Consider that the ideal $\widehat{\bm{U}}, \widehat{\bm{V}}$ are subject to unitary constraints, we formulate the overall training process as an optimization problem as follows:
\begin{equation}
\begin{aligned}
\min_{\bmmc{G}} \quad & \ell(\widehat{\bmmc{G}}), \\
\textrm{s.t.} \quad & \widehat{\bm{U}} \in \bmmc{P}, \\
                    & \widehat{\bm{V}^*} \in \bmmc{P}, \\
\end{aligned}
\label{eqn:optim-obj}
\end{equation}
where $\bmmc{P}$ denotes the unitary matrices set. Note that during the optimization, $\widehat{\bm{U}}, \widehat{\bm{V}^*}$ can be utilized in their format without recovering to $\widehat{\bmmc{G}}$. For simplicity, we still consistently use $\bmmc{G}$/$\widehat{\bmmc{G}}$ to denote $\{{\bm{U}, \bm{S}, \bm{V}^*}\}$/$\{\widehat{\bm{U}}, \bm{S}, \widehat{\bm{V}^*}\}$ in this section.

Performing direct optimization of the Eq. (\ref{eqn:optim-obj}) is challenging given the non-differentiable constraint $\bmmc{P}$. To solve this problem, we leverage the alternating direction optimization method that introduces an auxiliary variable ${\bm{\mathcal{Z}}}$. The Eq. \ref{eqn:optim-obj} can be then reformulated as follows:
\begin{equation}
\label{eq:optim-rew}
\begin{aligned}
\min_{\bmmc{G},\bmmc{Z}} \ell(\widehat{\bmmc{G}}) + g(\bmmc{Z}), ~~ \text{s.t.} ~~ \widehat{\bmmc{G}}=\bmmc{Z},
\end{aligned}
\end{equation}
where $g(\cdot)$ is an indicator function that takes a value of zero when the input variable corresponds to a unitary matrix; otherwise positive infinite. The corresponding augmented Lagrangian in the scaled dual form of Eq. \ref{eq:optim-rew} is then formulated as:
\begin{equation}
\begin{aligned}
\mathcal{L}_{\rho}(\bmmc{G}, \bmmc{Z}, \bmmc{U}) = &\ell(\widehat{\bmmc{G}}) + g(\bmmc{Z})\\
&+ \frac{\rho}{2} \left \| \widehat{\bmmc{G}} - \bmmc{Z} + \bmmc{U}\right \|_{F}^{2} + \frac{\rho}{2}\|\bmmc{U}\|_F^2,
\end{aligned}
\label{eqn:lagran}
\end{equation}
where $\bmmc{U}$ is the dual multiplier, and $\rho>0$ is the penalty parameter. The iterative update of these variables can be explicitly formulated as: 
\begin{equation}
\begin{aligned}
\bmmc{G}^{t+1}& = \argmin_{\bmmc{G}}~~\mathcal{L}_{\rho}\left ( \widehat{\bmmc{G}}, \bmmc{Z}^{t}, \bmmc{U}^{t} \right ),
\label{eqn:sub_w} 
\end{aligned}
\end{equation}
\begin{equation}
\begin{aligned}
\bmmc{Z}^{t+1}& = \argmin_{\bmmc{Z}}~~\mathcal{L}_{\rho}\left ( \widehat{\bmmc{G}}^{t+1}, \bmmc{Z}, \bmmc{U}^{t} \right ), 
\label{eqn:sub_z} 
\end{aligned}
\end{equation}
\begin{equation}
\begin{aligned}
\bmmc{U}^{t+1}& = \bmmc{U}^{t} + \widehat{\bmmc{G}}^{t+1} - \bmmc{Z}^{t+1}, \label{eqn:sub_u}
\end{aligned}
\end{equation}
where $t$ is the iteration step. Here, the original optimization problem (\ref{eqn:optim-obj}) is decomposed into two sub-problems (\ref{eqn:sub_w}) and (\ref{eqn:sub_z}), which can be solved independently as follows.

\textbf{Update $\bmmc{G}$ with standard optimizer.}
TT-cores $\bmmc{G}$ can be updated by the standard optimizer (e.g., SGD) with learning rate $\alpha$ as:
\begin{equation}
\begin{aligned}
\bmmc{G}\leftarrow\bmmc{G}-\alpha[\nabla_{\bmmc{G}} \ell(\widehat{\bmmc{G}})+\rho(\widehat{\bmmc{G}}-{\bmmc{Z}}+\bmmc{U})],
\label{eq:update_g}\\
\end{aligned}
\end{equation}

\textbf{Update $\bmmc{Z}$ with TT-unitary projection.} To update the introduced $\bmmc{Z}$, the optimization objective is:
\begin{equation}
\label{eq:updatez}
\min_{\bmmc{Z}\in\mathcal{P}}~\frac{\rho}{2}\|\widehat{\bmmc{G}}-\bmmc{Z}+\bmmc{U}\|_F^2.
\end{equation}
Given that $\bmmc{Z}$ is strictly constrained in the unitary matrix set $\bm{\mathcal{P}}$, the desired update can be executed through TT-unitary projection:
\begin{equation}
    \bmmc{Z}\leftarrow\textbf{TT-unitary}_{\mathcal{P}}(\widehat{\bmmc{G}}+\bmmc{U}).
\end{equation}
Specifically, the TT-unitary projection is defined as follows:
\begin{equation}
\label{eq:update_uni}
\begin{aligned}
\bm{T_U},\bm{S_U},\bm{R^*_{U}} =& \texttt{SVD}(\widehat{\bm{U}}), \\
\bm{U_p} =& \bm{T_U} \bm{R_U^*}, \\
\bm{T_V},\bm{S_V},\bm{R^*_{V}} =& \texttt{SVD}(\widehat{\bm{V^*}}), \\
\bm{V_p^*}  =& \bm{T_V} \bm{R_V^*}, \\
\bm{Z}  =& \bm{U_p} \bm{V_p^*}. \\
\end{aligned}
\end{equation}

\textbf{Update multiplier $\bmmc{U}$.} The dual multiplier $\bmmc{U}$ is updated as:
\begin{align}
\bmmc{U}\leftarrow \bmmc{U}+\widehat{\bmmc{G}}-\bmmc{Z}.
\label{eq:update_uv}
\end{align}

Upon completing the full update, the TT-cores $\bmmc{G}$ are fine-tuned with a standard optimizer (e.g., SGD or Adam) via robust training incorporating a noise and unitary forward process. The entire procedure is outlined in Algorithm \ref{alg:overall}.

\subsection{Reconfigurable Reshaping with Switch Matrices}

\begin{figure}[t]
	\centering 
	\includegraphics[width=1\linewidth]{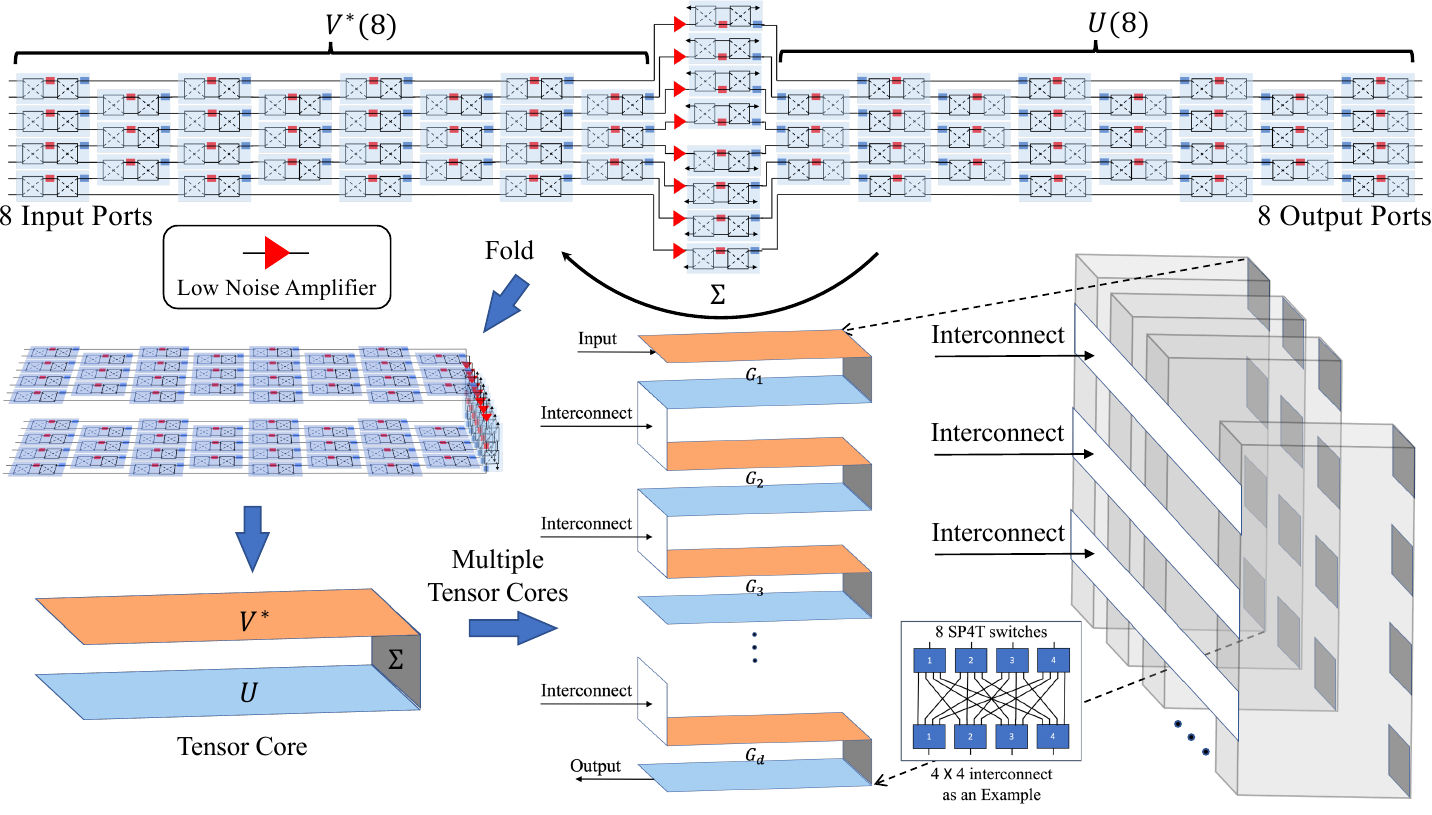}
	\caption{3D parallel stack realization for hardware space multiplexing of the tensor train.}
	\label{fig:Stack}
 \vspace{-2mm}
\end{figure}

When implementing a tensor core with RF interferometers, a tunable low-noise amplifier can be used on each channel to realize different diagonal values as well as power compensation to the propagation loss. Each tensor core can also be folded and stacked up for the area-efficient purpose, multiple stacks then can easily be paralleled for space multiplexing, as is shown in Fig. \ref{fig:Stack}. Frequency division multiplexing (FDM) combined with space-frequency interconnect solution \cite{xiao2021large} can also be applied to further speed up the calculation. Due to the reshaping process in the tensor-train method, interconnect layers between successive tensor cores are required. A feasible reconfigurable interconnect layer could not only enable different application implementations on the same hardware but also help to determine the best decomposition of the original matrix during in-situ training. Here for the first time, we propose that the tensor reshaping process during analog computation can be realized with a reconfigurable ${N\times N}$ RF switch matrices \cite{daneshmand2011rf}. Switch matrices are a conventional technique widely used for experiment automation. Recently, compact scalable RF switch matrix designs based on MEMS have been proposed and tested, which can handle RF frequencies up to 40GHz while every ${4 \times 4}$ switch matrix is only ${4mm^2}$ \cite{4956996}. For relatively small switch matrices in an FDM-enabled RF tensor train mesh, one can also build switch matrices with Single-Pole N-Throw (SPNT) switches and fixed interconnect wires, which can provide even better isolation. In addition, on the PCB platform, the relatively large device size makes reconfigurable interconnect between successor tensor cores manually with coaxial cables also a feasible yet straightforward solution during the test phase. 

\begin{algorithm}[t]
    \caption{Overall algorithm for TT-RFNN and RTT-RFNN.}
    \label{alg:overall}
    \textbf{Input:} Original RFNN $\bm{\mathcal{W}}$, target TT-ranks $\{r^*_i\}_{i=0}^{2d}$, noise budget $\lambda$, training epochs $T$.\par
    \textbf{Output:} TT-RFNN, RTT-RFNN with TT-cores $\{\bmmc{G}\}$.\par
    \begin{algorithmic}[1]
    
    \STATE $\bmmc{G}\leftarrow\textbf{TT-Decompose}(\bmmc{W}, \bm{r}^*)$;
    \STATE Obtain TT-RFNN after fine-tuning $\{\bmmc{G}\}$.
    \STATE Initialize $\bmmc{Z}$, $\bmmc{U}$;
    \FOR{$t=1$ \textbf{to} $T$}
    \STATE Update $\bmmc{U}$ using Eq. \ref{eq:update_uv};
    \STATE Update $\bmmc{G}$ using Eq. \ref{eq:update_g};
    \STATE Update $\bmmc{Z}$ using Eq. \ref{eq:update_uni};    
    \ENDFOR
    \STATE Obtain RTT-RFNN by fine-tuning $\{\bmmc{G}\}$.
    \end{algorithmic} 
\end{algorithm}

\subsection{Hardware Performance Estimation}
\label{hardware}

Given reshaped tensor core matrices ${\bm{G}\in\mathbb{R}^{M\times N}}$ with an input $\bm{x} \in \mathbb{R}^{N}$, the number of the RF interferometers required is $M(M-1)/2+N(N-1)/2+\texttt{max}(M, N)$ and $\texttt{max}(M, N)$ optional amplifiers. Here we suppose $M=N=k$ for analysis simplicity and the total number of RF interferometers is $k^2$, which is the same as its dimension of freedom. Since the speed of the electromagnetic (EM) wave on the device is of the same order as the speed of light and each device length is rough of its wavelength, therefore, the delay caused by each RF interferometer is one frequency period. We can estimate the delay of the tensor core ${{\tau}_G}$ is ${(4k-5)/f}$ for the triangular mesh and ${(2k+1)/f}$ for the rectangular mesh, where ${f}$ is the frequency of the EM wave. Since the rectangular mesh has the advantage of compact size, smaller delay, and better-balanced loss \cite{clements2016optimal}, the following estimation will focus on the rectangular mesh if not explicitly stated. Since the delay is a function of ${1/f}$, the higher frequency one chooses, a shorter device mesh and smaller delay can be achieved. For a typical radio frequency of several gigahertz, one can easily achieve nanoseconds order of delay. There are also interconnection layers between each tensor core, which are formed of intertwined transmission lines, the maximum delay of which can be ${N/2f}$ because for the worst case the length will be ${N}$ RF interferometer width ${N{\lambda}/2}$. Here we consider each RF interferometer has a height of ${{\lambda}/2}$, a rough estimation based on Fig.\ref{fig:RF_unit}(b) and its center frequency is 2GHz. Therefore, if the original weight matrix is ${\bm{W}\in\mathbb{R}^{N\times N}}$ and factorized into ${N/k}$ tensor trains each with ${d}$ tensor cores of dimension ${k \times k}$, the total number of RF interferometers required is ${dkN}$, with an area of ${dkN{\lambda}^2/2}$. If we take the average length for each interconnect layer, the ${(d-1)}$ interconnect layers cost a delay of ${dN/(4f)}$ and the ${d}$ tensor cores cost a delay of ${d{\tau}_G}$. Typically, ${k}$ will be much smaller than ${N}$, therefore, the entire delay is mostly contributed by all interconnect layers. For comparison, a ${N \times N}$ matrix without using the tensor-train technique requires ${N^2}$ RF interferometers, with an area of ${({\lambda}N)^2/2}$ and a delay of ${(2N+1)/f}$. It is worth mentioning that, the ${N/k}$ duplicated tensor trains we used in the estimation can be further reduced using the FDM method. Likewise, as another extreme case, one can also repeatedly utilize a single tensor train to perform only the forward analog computation on each core and memorize all outputs and perform tensor reshaping later on a digital computer, which will cost us the least RF interferometers, i.e., ${dk^2}$ units. The delay will be:
\begin{equation}
\begin{aligned}
\left[\frac{N}{k}\left(\tau_G+\tau_m\right)+\tau_r\right]\left(d-1\right),
\end{aligned}
\label{eq:delay1}
\end{equation}
\begin{equation}
\begin{aligned}
=2(d-1)N/f+\tau_{pc},
\end{aligned}
\label{eq:delay2}
\end{equation}
where ${\tau_{pc}=(N\tau_m/k+\tau_r)(d-1)}$ is the total delay cost by the computer, in which ${\tau_m}$ is the delay of memory accessing and ${\tau_r}$ is the delay of reshaping.

The tensor-train mesh requires much fewer RF interferometers and a shorter propagation length, which results in fewer active phase shifters and less power loss in propagation. It is worth mentioning that, each continuous thermo-optical phase modulator consumes about ${10mW}$ \cite{shen2017deep} while each discrete phase shifter on PCB platform \cite{10186076} consumes only ${0.24mW}$. We assume a typical insertion loss of transmission line on PCB to be ${0.25}$dB per wavelength, therefore, the total power consumption is dominated by all phase shifters, when the propagation length roughly ${<100}$ wavelengths. In addition, there are low loss PCB boards, such as Sheldahl Comclad XFx10mil \cite{1393206}, which has a measured loss tangent of 0.00024 while the typical value of a high-frequency PCB board is 0.004.

For a passive design, assuming the RF power detection rate is ${f_d\approx10}$MHz, which corresponds to ${10^7}$ N-dimensional analog matrix-vector multiplication in one second and is ${2N^2\times10^7}$ FLOPs per second. It's obvious that larger N and faster power detectors result in higher FLOPs per second. Let's assume ${N=100}$, therefore, the passive analog processor can perform 0.2 TFLOPs per second. Assume an RF power detector has a sensitivity of -60 dBm, the propagation loss of the original ${100 \times 100}$ device mesh will be 50dB, and the total input power necessary during forward propagation is estimated to be around ${P_{original}=10mW}$. However, Due to a much smaller propagation length, the propagation loss of a fully paralleled tensor train mesh will be greatly reduced to 8dB (assuming ${k=4}$, ${d=4}$ and neglect the propagation loss from low loss interconnection wires), therefore, its input power can be reduced to ${P_{parallel}=6.3{\times}10^{-4}mW}$. In general, ${dk{\ll}N}$, one can derive that ${P_{original}/P_{parallel}=10^{(N-dk)/20}{\gg}1}$, which proves that the tensor train solutions can greatly reduce the power consumption.

\begin{table}[t]
\centering
\caption{Hardware performance estimation. ``Pow. Cons." denotes ``Power Consumption". }
\setlength{\tabcolsep}{1pt}
\scalebox{1}{
\begin{tabular}{c|ccc}
\toprule 
                                                                        & 
\multirow{2}{*}{\makecell{{RFNN}\\{(Original)}}} & 
\multirow{2}{*}{\makecell{{TT-RFNN}\\{(Parallel)}}} & 
\multirow{2}{*}{\makecell{{TT-RFNN}\\{(Single})}} \\
 & & \\
\midrule
\# RF interferometers                    &       $N^2$          &      $dkN$              &         $dk^2$          \\
Delay                          &         $(2N+1)/f$         &      $dN/4f$       &            $2(d-1)N/f+\tau_{pc}$    \\
Area                           &          $N^2\lambda^2/2$        &        $dkN\lambda^2/2$             &         $dk^2\lambda^2/2$          \\
Active Pow. Cons.       &        $N^2P_{ps}$          &          $dkNP_{ps}$           &         $dkNP_{ps}+P_{pc}$          \\
Passive Pow. Cons.${^*}$ &          $N10^{\frac{0.5N-60}{10}}$       &            $N10^{\frac{0.5dk-60}{10}}$         &          $N10^{\frac{0.5dk-60}{10}}+P_{pc}$         \\
\bottomrule
\end{tabular}
}
\label{figure:hw} 
\vspace{1mm}
\parbox{0.9\linewidth}{
\footnotesize{ ${P_{ps}}$: Power consumption of a single phase shifter; ${P_{pc}}$: Power consumption related to the computer.

$^*$ Unit is ${mW}$ under assumption of propagation loss 0.25dB per wavelength and detector sensitivity -60dBm. }}\\
\vspace{-4mm}
\end{table}


\section{Experiments}

\begin{table*}[ht]
\centering
\caption{Results for TT-RFNN of MLP-1 on the MNIST dataset and VGG-16 on the CIFAR-10 dataset. }
\setlength{\tabcolsep}{2.5pt}
\scalebox{1}{
\begin{tabular}{ccccccccccc}
\toprule
& & \textbf{Rank Setting} & \textbf{Method} & \textbf{Top-1 Accuracy (\%)} & \textbf{\#Params} & \textbf{Compression Ratio}  & \textbf{\#PS}  & \textbf{\#RF interferometers} & \textbf{Area $\downarrow$} \\
\hline
\multirow{10}{*}{\textbf{MLP-1}} & & & Original & 98.3 & 1.86M & 1$\times$ & 4.80M & 2.40M & 1$\times$ \\
\hline 
& \multirow{3}{*}{\textbf{Shape-1}} & Rank-1-1 & TT-RFNN (Ours) & 96.9 & 9.53K & 193.6$\times$ & 55.2K & 27.6K & 86.9$\times$\\
& & Rank-1-2 & TT-RFNN (Ours) & 97.5 & 16.3K & 112.9$\times$ & 97.2K & 48.6K & 49.4$\times$ \\
& & Rank-1-3 & TT-RFNN (Ours) & 97.8 & 24.8K & 74.4$\times$ & 151.8K & 75.9K & 31.7$\times$ \\
\cmidrule(lr){2-10}
& \multirow{3}{*}{\textbf{Shape-2}} & Rank-2-1 & TT-RFNN (Ours) & 97.0 & 9.74K & 189.4$\times$ & 55.6K & 27.8K & 86.3$\times$\\
& & Rank-2-2 & TT-RFNN (Ours) & 97.6 & 16.4K & 112.4$\times$ & 97.4K & 48.7K & 49.3$\times$ \\
& & Rank-2-3 & TT-RFNN (Ours) & 98.0 & 24.6K & 75.1$\times$ & 152.0K & 76.0K & 31.6$\times$ \\
\midrule
\multirow{10}{*}{\textbf{VGG16-BN}} &  & & Original & 94.19 & 15.25M & 1$\times$ & 128.4M & 64.2M & 1$\times$ \\
\hline 
& \multirow{3}{*}{\textbf{Shape-1}} & Rank-3-1 & TT-RFNN (Ours) & 93.78 & 2.39M & 6.38$\times$ & 27.2M & 13.6M & 4.72$\times$ \\
& & Rank-3-2 & TT-RFNN (Ours) & 93.66 & 2.45M & 6.21$\times$ & 28.6M & 14.3M & 4.48$\times$ \\
& & Rank-3-3 & TT-RFNN (Ours) & 94.08 & 2.88M & 5.28$\times$ & 41.6M & 20.8M & 3.08$\times$ \\
\cmidrule(lr){2-10}
& \multirow{3}{*}{\textbf{Shape-2}} & Rank-4-1 & TT-RFNN (Ours) & 93.94 & 3.67M & 4.15$\times$ & 22.6M & 11.3M & 5.65$\times$ \\
& & Rank-4-2 & TT-RFNN (Ours) & 94.12 & 3.77M & 4.04$\times$ & 23.8M & 11.9M & 5.41$\times$ \\
& & Rank-4-3 & TT-RFNN (Ours) & 94.13 & 3.96M & 3.84$\times$ & 24.8M & 12.4M & 5.16$\times$ \\
\bottomrule
\end{tabular}
}
\label{table:tt-rfnn}
\end{table*}

 \begin{figure*}[ht] 
 \centering
  \subfigure[Power saving with various rank settings on MLP-1. A larger value is better.]
     {\includegraphics[width=0.35\textwidth]{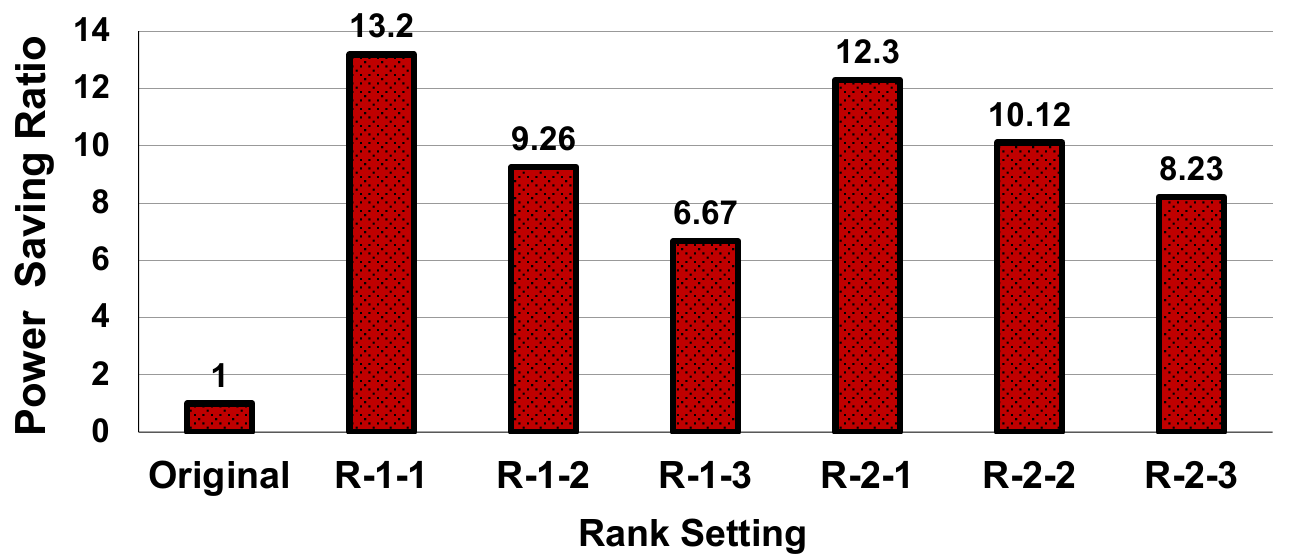}
     } 
 \subfigure[Accuracy vs. compression ratio on MLP-1.]
     {\includegraphics[width=0.3\textwidth]{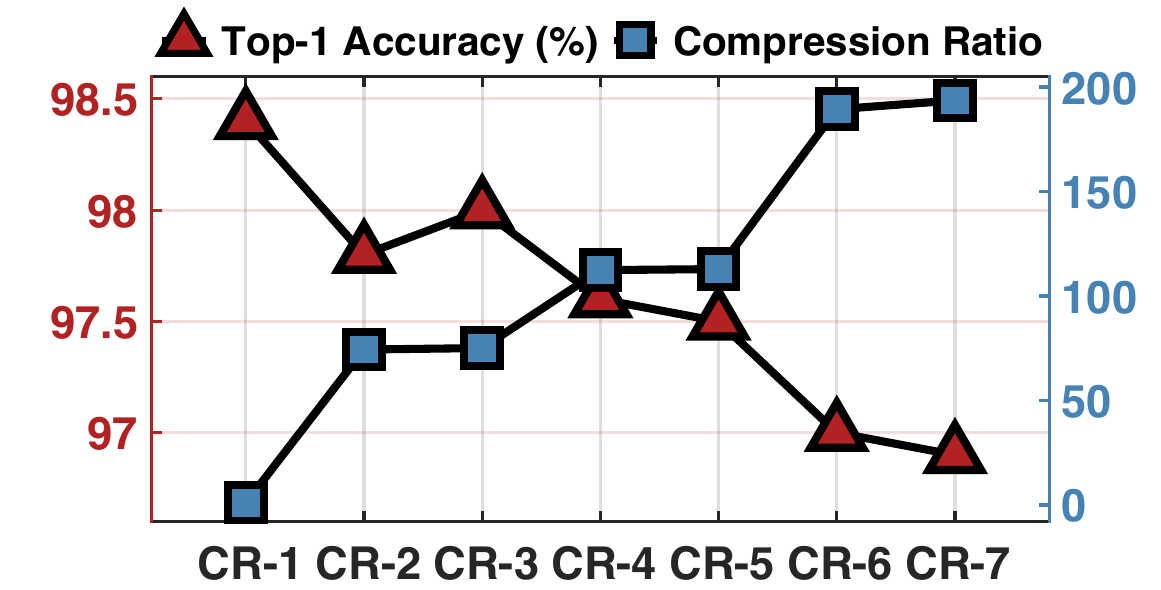}
     } 
 \subfigure[Accuracy vs. compression ratio on VGG16-BN.]
 {\includegraphics[width=0.3\textwidth]{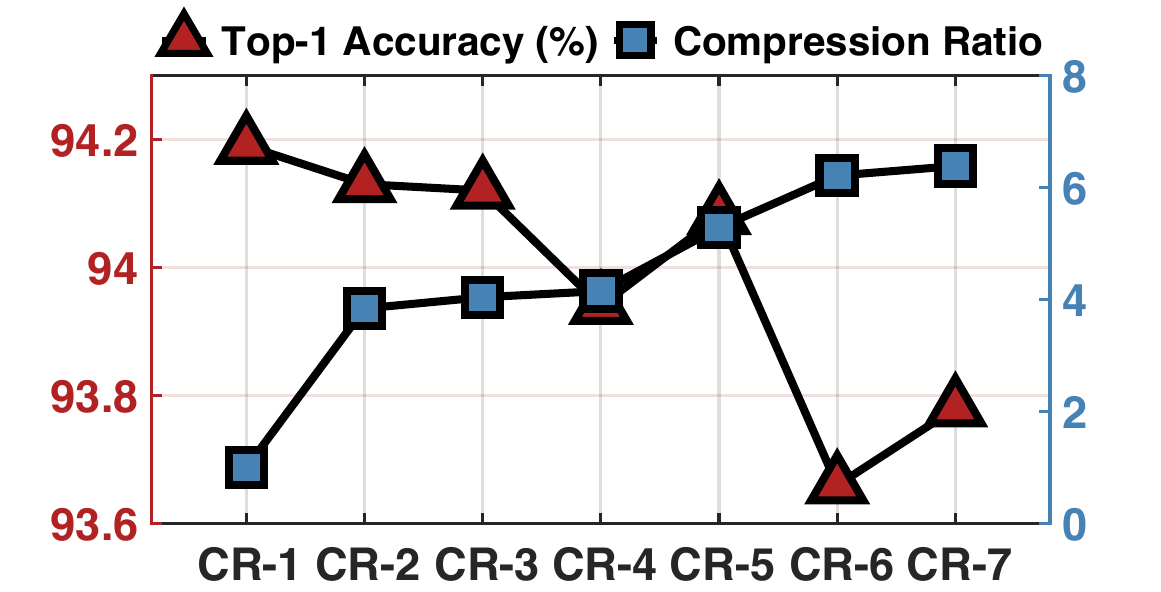}
     } 
         \vspace{-2mm}
    \caption{Comparison of the power saving ratio, top-1 accuracy vs. compression ratio.}
    \label{fig:trend}
    \vspace{-2mm}
\end{figure*}

\textbf{Datasets.} We evaluate the proposed TT-RFNN and the R-TT-RFNN on MNIST \cite{lecun1998mnist} and CIFAR-10 \cite{krizhevsky2009learning} datasets. On the CIFAR-10 dataset, we adopt basic augmentation, e.g., random crop and flip. Images in MNIST and CIFAR-10 are with (Height$\times$Width$\times$Channel) of (28$\times$28$\times$1) and (32$\times$32$\times$3), respectively. 

\textbf{Models.} For the MNIST dataset, we use two different MLP models. MLP-1 (784-1024-1024-10) has three fully-connected layers for evaluating TT-RFNN, and MLP-2 (784-1024-10) has two fully-connected layers for RTT-RFNN, where 784/1024/10 denotes the number of neurons. For the CIFAR-10 dataset, VGG16-BN is adopted as our model. 

\textbf{Shapes, Ranks, and RF interferometers.} Table \ref{result:shape_rank} shows some example settings of the shapes and ranks used in our experiments. In Table \ref{table:tt-rfnn}, the number of RF interferometers and area are calculated without considering parallel computing. Disregarding the RF interferometers in the amplifier units, the minimum number of RF interferometers for a weight matrix/tensor core matrix $\bm{W} \in \mathbb{R}^{M \times N}$ is $M(M-1)/2 + N(N-1)/2$. Power consumption is estimated only with active power consumption of phase shifters.


\textbf{Other Hyperparameters.} For training MNIST, we use Adam as the optimizer with the learning rate 1e-3. For training CIFAR-10, we use SGD with a learning rate of 0.05, momentum of 0.9, and weight decay of 5e-4. All models are implemented with PyTorch framework \cite{paszke2019pytorch}.

\subsection{Efficiency of TT-RFNN}


\begin{table}[t]
\centering
\caption{Examples of ranks of TT-RFNN in Table \ref{table:tt-rfnn}.}
\setlength{\tabcolsep}{15pt}
\scalebox{0.8}{
\begin{tabular}{ccc}
\toprule
 & \textbf{Layer} & \textbf{Ranks} \\
\midrule
\multicolumn{3}{c}{MLP-1, Shape-2} \\
\hline
\multirow{3}{*}{Rank-2-1} & FC-1 & [1, 8, 12, 12, 12, 12, 12, 7, 1] \\
& FC-2 & [1, 8, 12, 12, 12, 12, 12, 8, 1] \\
& FC-3 & [1, 5, 5, 10, 8, 8, 8, 8, 1] \\
\hline
\multirow{3}{*}{Rank-2-2} & FC-1 & [1, 8, 16, 16, 16, 16, 16, 7, 1] \\
& FC-2 & [1, 8, 16, 16, 16, 16, 16, 8, 1] \\
& FC-3 & [1, 5, 5, 10, 12, 12, 12, 8, 1] \\
\hline
\multirow{3}{*}{Rank-2-3} & FC-1 & [1, 8, 20, 20, 20, 20, 20, 7, 1] \\
& FC-2 & [1, 8, 20, 20, 20, 20, 20, 8, 1] \\
& FC-3 & [1, 5, 5, 10, 16, 16, 16, 8, 1] \\
\midrule
\multicolumn{3}{c}{VGG16-BN, Shape-2} \\
\hline
\multirow{5}{*}{Rank-4-1} & Feature-27 & [1, 8, 32, 32, 48, 48, 32, 32, 8, 1] \\
& Feature-30 & [1, 8, 32, 32, 48, 48, 32, 32, 8, 1] \\
& Feature-34 & [1, 8, 32, 32, 48, 48, 32, 32, 8, 1] \\
& Feature-37 & [1, 8, 32, 32, 48, 48, 32, 32, 8, 1] \\
& Feature-40 & [1, 8, 32, 32, 48, 48, 32, 32, 8, 1] \\
\hline
\multirow{5}{*}{Rank-4-2} & Feature-27 & [1, 8, 32, 32, 64, 64, 32, 32, 8, 1] \\
& Feature-30 & [1, 8, 32, 32, 64, 64, 32, 32, 8, 1] \\
& Feature-34 & [1, 8, 32, 32, 64, 64, 32, 32, 8, 1] \\
& Feature-37 & [1, 8, 32, 32, 64, 64, 32, 32, 8, 1] \\
& Feature-40 & [1, 8, 32, 32, 64, 64, 32, 32, 8, 1] \\
\hline
\multirow{5}{*}{Rank-4-3} & Feature-27 & [1, 8, 32, 64, 72, 72, 64, 32, 8, 1] \\
& Feature-30 & [1, 8, 32, 64, 72, 72, 64, 32, 8, 1] \\
& Feature-34 & [1, 8, 32, 64, 72, 72, 64, 32, 8, 1] \\
& Feature-37 & [1, 8, 32, 64, 72, 72, 64, 32, 8, 1] \\
& Feature-40 & [1, 8, 32, 64, 72, 72, 64, 32, 8, 1] \\
\bottomrule
\end{tabular}
}
\label{result:shape_rank}
\vspace{-4mm}
\end{table}

\textbf{MNIST dataset.} In the context of MLP-1 with architecture 784-1024-1024-10, we evaluate two different decomposed shapes. For each shape, three rank settings are used to demonstrate the efficiency of our proposed method. With an original MLP-1 with a top-1 accuracy of 98.3$\%$, we apply our approach to derive a compact TT-RFNN. Table \ref{table:tt-rfnn} and Fig. \ref{fig:trend} demonstrate the effectiveness of TT-RFNN. For instance, for a pre-trained model with 1.86M parameters, our method can significantly compress the model to 9.53K parameters when utilizing Shape-1 in which ``FC-1" [4, 8, 8, 4, 4, 7, 7, 4], ``FC-2" [4, 8, 8, 4, 4, 8, 8, 4] and ``FC-3" [1, 5, 2, 1, 4, 8, 8, 4]. This amounts to approximately 194$\times$ reduction in storage complexity under the Rank-1-1 configuration. Meanwhile, our approach brings $87\times$ reduction in the usage of RF interferometers, achieving substantial area savings while maintaining an accuracy close to 97$\%$. Meanwhile, it saves $13.2\times$ power consumption compared to the original model. Experiments using Rank-1-2 and Rank-1-3 settings offer a relatively lighter compression ratio, yielding about $113\times$ and $74\times$ respectively. Simultaneously, they achieve a power consumption reduction of $9.26\times$ and $6.67\times$ compared to the original model. Such high compression ratios only slightly affect the top-1 accuracy of $0.8\%$ and $0.5\%$ degradation. Similarly, when adopting Shape-2 with ``FC-1" [8, 4, 4, 8, 7, 4, 4, 7], ``FC-2" [8, 4, 4, 8, 8, 4, 4, 8] and``FC-3" [5, 1, 1, 2, 8, 4, 4, 8], our approach also bring significant complexity reduction.



\textbf{CIFAR-10 dataset.} Utilizing the original VGG-16-BN model, we conduct an experiment with two tensor decomposition shapes. For each shape, we consider three rank settings. As shown in Table \ref{table:tt-rfnn}, for an original model with 15.25M parameters, when applied with Shape-1 that ``Feature-27, -30, -34, -37, -40" are all with shape [32, 16, 9, 16, 32] (Rank-3-1 setting), our approach can significantly compress the model significantly and bring its size down to 2.39M parameters and 4.72$\times$ area reduction while the model accuracy remains nearly at 93.78$\%$. Further experiments under Rank-3-2 and Rank-3-3 settings also show that under approximately $4.48\times$ and $3.08\times$ area reduction, the compressed models can achieve the accuracy of $93.66\%$ and $94.08\%$. In addition, experiments with Shape-2 that ``Feature-27, -30, -34, -37, -40" are all with shape [8, 4, 4, 4, 9, 4, 4, 4, 8] configuration also show good performance. 

\subsection{Robustness of TT-RFNN}

\begin{figure}[t] 
 \subfigure[$\lambda=0.01$.]
     {\includegraphics[width=0.235\textwidth]{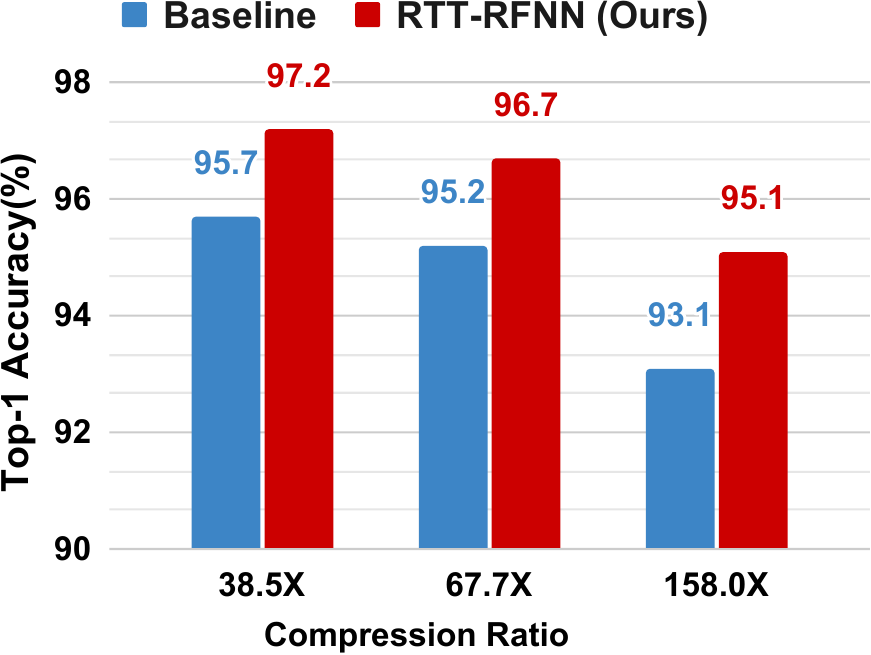}
     } 
 \subfigure[$\lambda=0.02$.]
 {\includegraphics[width=0.235\textwidth]{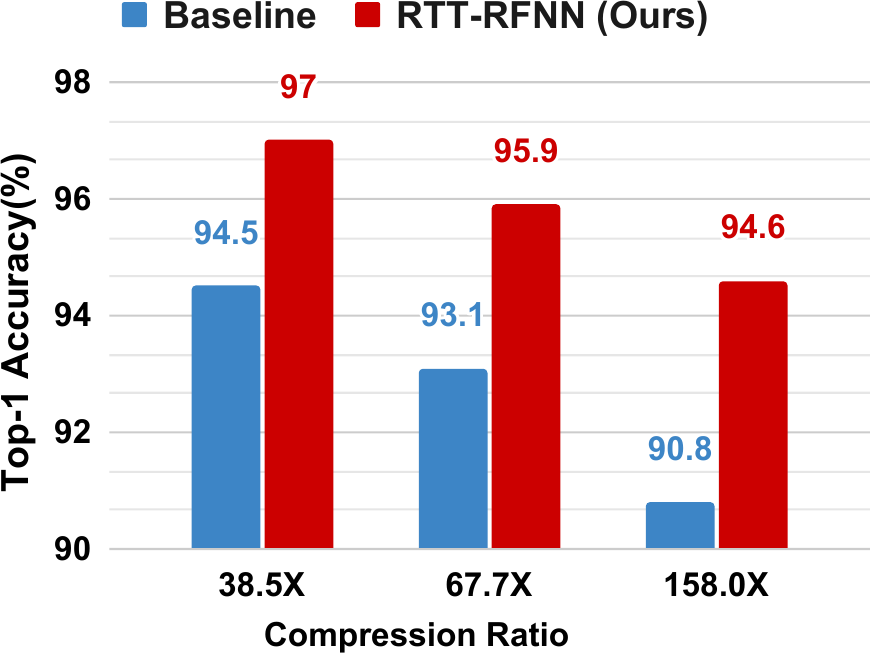}
     } 
    \caption{Results for RTT-RFNN of MLP-2 (784-1024-10) with Shape-1 setting on MNIST dataset.}
    \label{fig:robust-mlp2-s1}
\end{figure}

\begin{figure}[t] 
 \subfigure[$\lambda=0.01$.]
     {\includegraphics[width=0.235\textwidth]{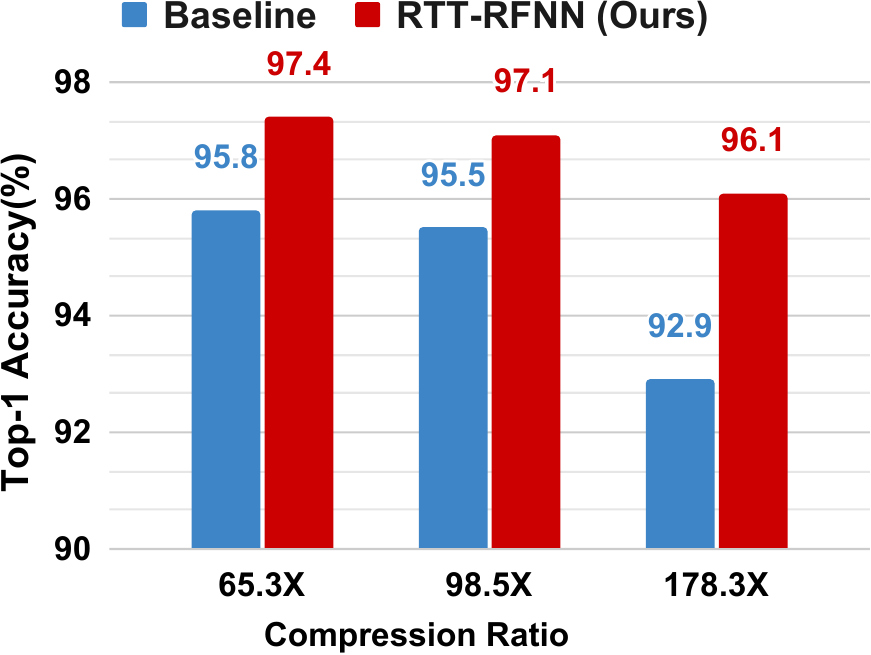}
     } 
 \subfigure[$\lambda=0.02$.]
 {\includegraphics[width=0.235\textwidth]{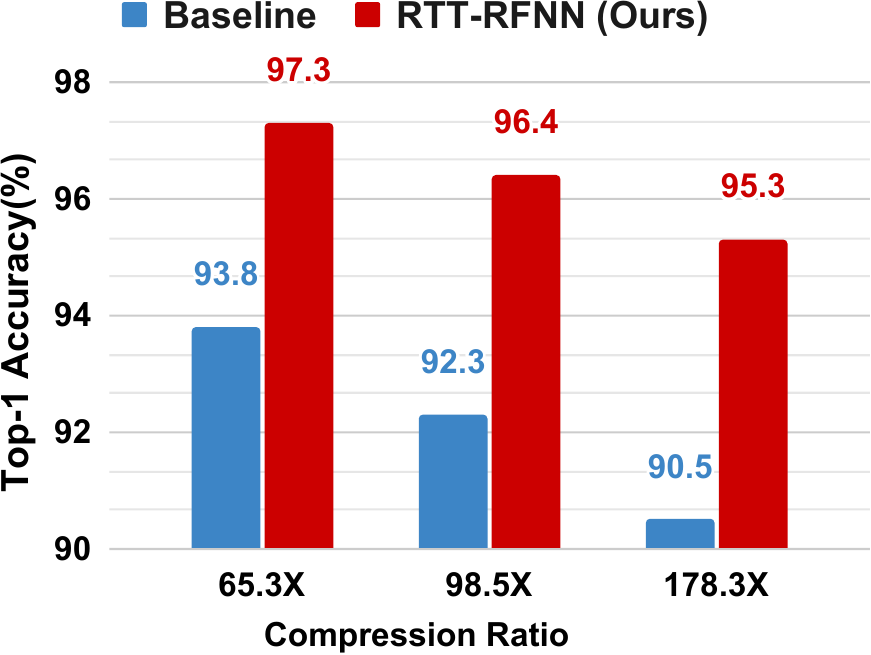}
     } 
    \caption{Results for RTT-RFNN of MLP-2 (784-1024-10) with Shape-2 setting on MNIST dataset.}
    \label{fig:robust-mlp2-s2}
    \vspace{-2mm}
\end{figure}

\begin{figure}[t] 
 \subfigure[$\lambda=0.04$.]
     {\includegraphics[width=0.235\textwidth]{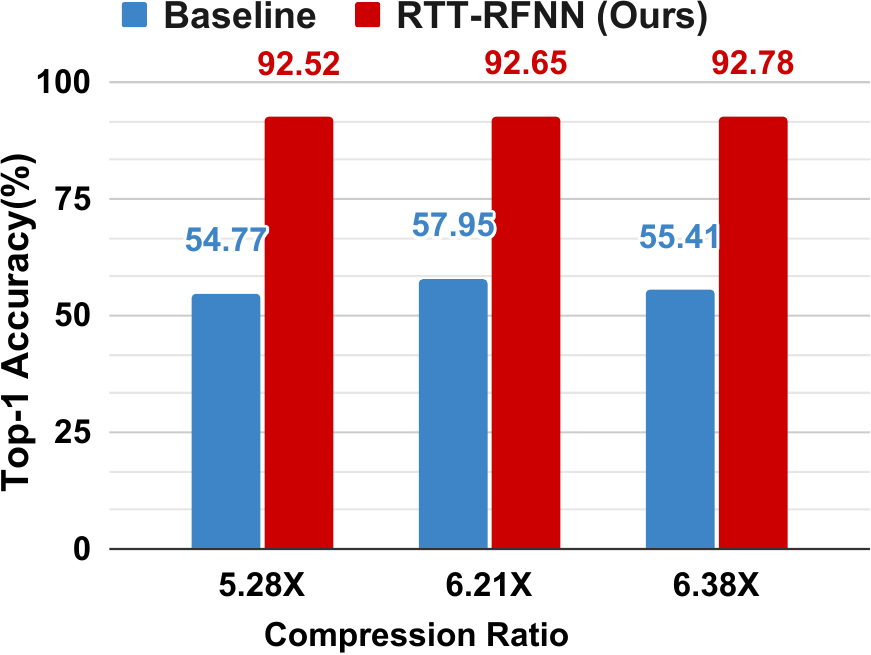}
     } 
 \subfigure[$\lambda=0.05$.]
 {\includegraphics[width=0.235\textwidth]{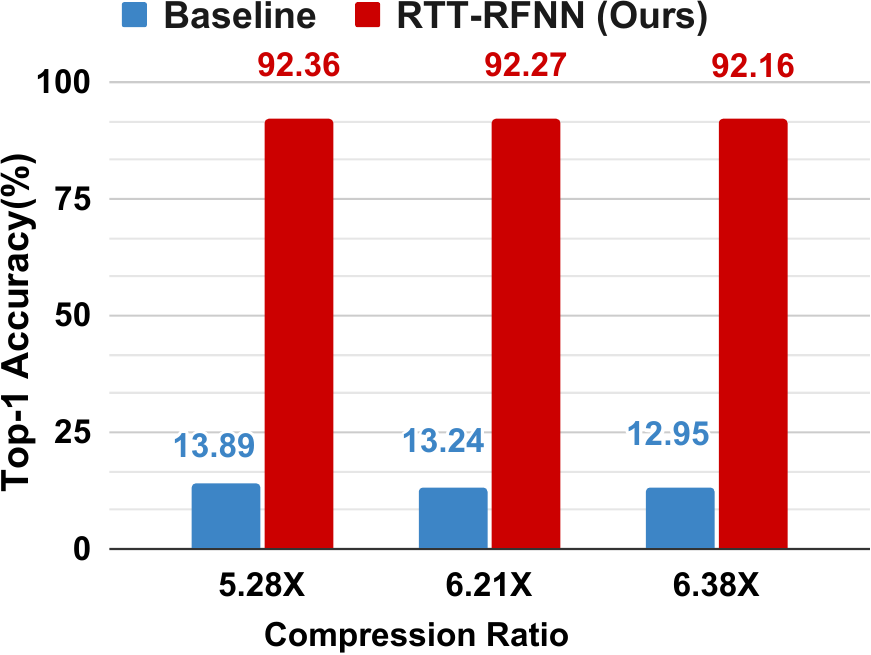}
     } 
    \caption{Results for RTT-RFNN of VGG16-BN with Shape-1 setting on CIFAR-10 dataset.}
    \label{fig:robust-vgg}
    \vspace{-2mm}
\end{figure}

To validate the effectiveness of our proposed approach, we train RTT-RFNN, compared with model TT-RFNN affected by random Gaussian noises, which serves as our baseline. The evaluation introduces various magnitudes $\lambda$ of Gaussian noise into the TT-RFNN. 

\textbf{MNIST dataset.} Given an original MLP-2 model (784-1024-10) with $97.8\%$ top-1 accuracy, under the compression ratio of 39$\times$ with the Shape-3 with ``FC-1" [4, 8, 8, 4, 4, 7, 7, 4], ``FC-2" [1, 5, 2, 1, 4, 8, 8, 4], the ideal TT-RFNN can achieve $97.8\%$ top-1 accuracy. However, considering the noise, the TT-RFNN would incur the performance degradation to $95.7\%$ under the $\lambda=0.01$. Our proposed method allows the accuracy to be sustained at $97.2\%$ under the $39\times$ compression ratio, which increases 1.5$\%$ than the vulnerable TT-RFNN under the same noise condition. Meanwhile, RTT-RFNN only exhibits a $0.6\%$ accuracy drop compared to the noise-free TT-RFNN. 

\textbf{CIFAR-10 dataset.} Considering an original VGG16-BN model with $94.19\%$ top-1 accuracy with the decomposed Shape-1. With the compression ratio of 6.38$\times$, the TT-RFNN can achieve $93.78\%$ top-1 accuracy without noise effect. When considering the practical mapping noise, the performance of the noise TT-RFNN diminishes to $55.41\%$ under the $\lambda=0.04$. By using our method, RTT-RFNN can keep the accuracy as $92.78\%$ under the $6.38\times$ compression ratio, which increases about 37$\%$ than the TT-RFNN under the same level of noise. The TT-RFNN would incur the performance degradation to $12.95\%$ under the $\lambda=0.05$. However, RTT-RFNN can keep the accuracy as $92.16\%$ under the $6.38\times$ compression ratio, which increases about 79$\%$ than the TT-RFNN under the same level of noise. Empirical results from experiments demonstrate that our approach can consistently yield efficient and robust models.

\section{Conclusion}

In this paper, we introduce a method designed to produce an efficient and robust RFNN capable of both reducing the count of RF interferometers and mitigating deployment noise from a general perspective. Specifically, our method first generates a TT-RFNN in which the individual layer consists of a series of third-order TT-cores, resulting in a remarkable reduction of parameters and a considerable saving in RF interferometer count and energy consumption. Subsequently, to address the noise intrinsically associated with RF device implementations, we utilize a robustness solver to develop the RTT-RFNN. To flexibly adapt RTT-RFNN to diverse reshape operations, we also provide a reconfigurable solution based on switch matrices. Empirical evaluations conducted on the MNIST and CIFAR-10 datasets show the efficacy of our proposed approach.

\section*{Acknowledgment}
This work was supported in part by the Defense Advanced Research Projects Agency under Grant D19AP00030, in part by the Army Research Laboratory under Grant W911NF2120349 and in part by National Science Foundation Award CCF-1955909.

\bibliographystyle{IEEEtran}
\bibliography{IEEEfull}

\end{document}